\documentclass[10pt]{article}
\usepackage{physics}
\usepackage{qcircuit}
\usepackage{float}

\begin{document}

\title{Quantum Bit Error Avoidance}

\author{A.Y. Shiekh\footnote{\tt ashiekh@coloradomesa.edu} \\
	Department of Physics \\
	Colorado Mesa University \\
	Grand Junction, CO \\
	U.S.A.}

\date{}

\maketitle

\begin{abstract}
	Qubit errors might be avoided by using the quantum Zeno effect to inhibit evolution.
\end{abstract}

\section{Errors}
Errors need to be dealt with and there are two possible approaches to this problem, error correction and error avoidance.

\subsection{Digital versus analogue}
In digital systems only certain values are allowed, in contrast to analogue where all values in a range are valid. The advantage to digital is that errors may be detected and so corrected, but there are differences between quantum and classical digital systems.

\section{The one bit error assumption}

\subsection{Classical digital systems}
For the sake of a concrete example we may be thinking of voltage levels, where logical $0$ will be a certain low voltage range, and logical $1$ at a higher range for the classical bit (think TTL or CMOS). Now there will be voltage fluctuations around but so long as they are not excessive the chance of a single logical error is small and that of a double errors smaller still if they are independent. For such a system the assumption of a single bit error is valid, as the only error that can happen for a classical binary system is a bit flip error; this is not the case for a binary quantum system.

\subsection{Quantum digital systems}
Quantum states have both digital and analogue aspects and a quantum binary system can suffer from much more than just a bit flip error; start by looking at a qubit
\begin{equation}
\ket{\psi} = \alpha_0 \ket{0} + \alpha_1 \ket{1}
\end{equation}
the result of measurement is digital (zero or one) but the amplitudes and so probabilities are analogue, and errors may occur not just in the digital part (bit flip) but also in the analogue part (the amplitude and just the phase) \cite{NielsenChuang}.

If we look at a multi-qubit system, say
\begin{equation}
\ket{\psi} = \alpha_0 \ket{000} + \alpha_1 \ket{111}
\end{equation}
with a mind to error protection by majority vote, then it is unrealistic to assume that only one bit error is most likely. In reality all bits are likely to suffer an error, albeit small for all bits, and the original becomes
\begin{equation}
\begin{aligned}
	\alpha_{000} \ket{000} + \alpha_{111} \ket{111}& + \\
	\varepsilon_{001} \ket{001} + \varepsilon_{010} \ket{010}& + 
	\varepsilon_{011} \ket{011} + \varepsilon_{100} \ket{100} + 
	\varepsilon_{101} \ket{101} + \varepsilon_{110} \ket{110} 
\end{aligned}
\end{equation}
where $\varepsilon_{001}$ through $\varepsilon_{110}$ are small. Here classical and quantum digital systems differ, and error correction becomes problematic for the quantum system.

But a different approach is possible since quantum theory has a mechanism not available to classical systems, namely the quantum Zeno effect where the qubit can be decoupled from its environment so avoiding error in the presence of noise rather than correcting for it.

\section{Quantum Zeno effect}
Rapid repeated measurement can stop a quantum system from evolving \cite{MisraSudarshan}, a phenomenon that has been experimentally confirmed \cite{ItanoHeinzenBollingerWineland} and might be used to fight against decoherence; the problem is that measurement will cause a general quantum state to collapse and one needs to be a little more clever in implementing the quantum Zeno effect (also known as the Turing paradox).

For very short times we may assume the potential in Schr{\"o}dinger's equation \cite{Griffiths}
\begin{equation}
i \hbar \frac{\partial \Psi}{\partial t} = - \frac{\hbar^2}{2m} \frac{\partial^2 \Psi}{\partial x^2} + V \Psi
\end{equation}
is constant. We can then separate variables $\Psi(x,t) = \phi(x) \ \psi(t)$, to yield:
\begin{equation}
i \hbar \frac{d \psi}{d t} = E \psi
\end{equation}
where $E$ is the energy; the factor of $i$ in Schr{\"o}dinger's equation will play a very important role.

The evolution of the state for very short times must then be of the form
\begin{equation}
\psi(t) = \psi(0) \left( 1-i \frac{E}{\hbar} t + \mathcal{O}(t^2) \right)
\end{equation}
so if we keep repeating the measurement of an eigenstate after a short time ($t = T/n$) we get the probability ($| \psi |^2 $) of the original state as
\begin{equation}
\left( 1 - \mathcal{O}\left( \frac{1}{n^2} \right) + \ldots \right)^n
\end{equation}
Now in the limit of continuous observation
\begin{equation}
\lim\limits_{n \to \infty} \left( 1 - \frac{c}{n^2} \right)^n = 1
\end{equation}
the probability becomes unity, so we see that in quantum mechanics `The watched kettle never boils'.

\section{Qubit encoding and error avoidance}
The dilemma now is that the act of measurement of a non-eigenstate causes the quantum state to collapse, and so while the Zeno effect might be used to protect the state from the environment, it will disrupt that state in the process, so we need to use entanglement to our advantage.

Start with a single qubit
\begin{equation}
\ket{\psi} = \alpha_0 \ket{0} + \alpha_1 \ket{1}
\end{equation}
and entangle it with an auxiliary bit using the following quantum encoder circuit
 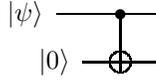
\begin{figure}[H]
$$
\Qcircuit @C=1em @R=1em @!R { 
	\lstick{\ket{\psi}}	
	& \qw 				& \ctrl{1} 	& \qw \\ 
	&\lstick{\ket{0}} 	& \targ 	& \qw
}
$$
\caption{Bit Encoding}
    \end{figure}
\noindent to yield
\begin{equation}
\alpha_0 \ket{00} + \alpha_1 \ket{11}
\end{equation}

Now first consider the case where the qubit has not picked up any error, and perform the following processing
 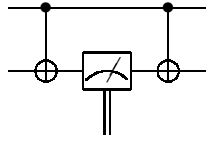
\begin{figure}[H]
$$
\Qcircuit @C=1em @R=1em @!R { 
	& \ctrl{1} 	& \qw 		& \ctrl{1} 	& \qw \\ 
	& \targ 	& \meter 	& \targ 	& \qw \\ 
	&			& \cwx[-1]	&			& 
}
$$
\caption{Zeno measurement}
\label{Zeno}
\end{figure}

In general, the act of measurement would disturb the quantum state, but let us follow the processed state and in particular the auxiliary bit.
\begin{equation}
\begin{aligned}
						& \alpha_0 \ket{00} + \alpha_1 \ket{11} \\
\rightarrow \hskip1em	& \alpha_0 \ket{00} + \alpha_1 \ket{10} = 
(\alpha_0 \ket{0} + \alpha_1 \ket{1}) \overbrace{\ket{0}}^\mathrm{aux} \\
\rightarrow \hskip1em	& \alpha_0 \ket{00} + \alpha_1 \ket{11}
\end{aligned}
\end{equation}
Note that the auxiliary bit is unaffected by the act of measurement.

We can now start over and consider the appearance of errors over a very short time duration; what should have been
\begin{equation}
\alpha_0 \ket{00} + \alpha_1 \ket{11}
\end{equation}
becomes
\begin{equation}
\begin{aligned}
\left( 1-i \frac{E_1}{\hbar} t + \mathcal{O}(t^2) \right) \left( 1-i \frac{E_2}{\hbar} t + \mathcal{O}(t^2) \right)
( \alpha_0 \ket{00} + \alpha_1 \ket{11} ) + \\
\left( \frac{E_1+E_2}{\hbar} t + \mathcal{O}(t^2) \right) 
( \alpha_{01} \ket{01} + \alpha_{10} \ket{10} )
\end{aligned}
\end{equation}
see the above discussion of the quantum Zeno effect. This is now processed through the error avoidance circuit (Fig:\ref{Zeno}) given above to first yield
\begin{equation}
\begin{aligned}
\left( 1-i \frac{E_1}{\hbar} t + \mathcal{O}(t^2) \right) \left( 1-i \frac{E_2}{\hbar} t + \mathcal{O}(t^2) \right) 
(\alpha_{0} \ket{0} + \alpha_{1} \ket{1} & \overbrace{\ket{0}}^\mathrm{aux} + \\
\left( \frac{E_1+E_2}{\hbar} t + \mathcal{O}(t^2) \right) 
(\alpha_{01} \ket{0} + \alpha_{10} \ket{1})	& \overbrace{\ket{1}}^\mathrm{aux}
\end{aligned}
\end{equation}
where a fast measurement on the auxiliary bit hinders evolution, yielding
\begin{equation}
(\alpha_{0} \ket{0} + \alpha_{1} \ket{1}) \ket{0}
\end{equation}
which is then restored to
\begin{equation}
\alpha_0 \ket{00} + \alpha_1 \ket{11}
\end{equation}
Error detection is present as a non-zero measured value of the auxiliary would indicate the process had failed.

In practice one would have more than one auxiliary bit measured alternately so while one auxiliary bit is being processed, the other is still in place.

\section*{Conclusion}
While it may not be possible to achieve quantum error correction, quantum error avoidance may be possible using the quantum Zeno effect to decouple the system from its environment.

\end{document}